\documentstyle[12pt]{article}
\begin{document}
\begin{flushright}
PRA-HEP-93/05\\
May 1993
\end{flushright}
\vspace{5ex}
\begin{center}
{\Large {\bf On parton distributions beyond the
leading order }} \\
\vspace{0.5cm}
{Ji\v{r}\'{\i} Ch\'{y}la \\
\vspace{0.4cm}
Institute of Physics, AV\v{C}R,
Na Slovance 2, Prague 8, Czech Republic\\}
\vspace{0.5cm}
{\bf Abstract}
\end{center}
The importance of properly taking into account the factorization
scheme dependence of parton distribution functions is emphasized.
A serious error in the usual handling of this topic is pointed
out and the correct procedure for transforming parton distribution
functions from one factorisation scheme to another recalled.
It is shown that the conventional $\overline{\rm {MS}}$ and DIS
definitions thereof are ill-defined due to the lack of distinction
between the factorisation scheme dependence of parton distribution
functions and renormalisation scheme dependence of the strong
coupling constant $\alpha_s$. A novel definition of parton
distribution functions is suggested and its role in the construction
of consistent next-to-leading order event generators briefly outlined.

\vspace{1.0cm}
\section{Introduction}
During recent years significant progress in the determination
of parton distribution functions (p.d.f.) in the nucleon has
been achieved,
basically as a result of new data \cite{CCFR,NMC}, combined with
more sophisticated and reliable theoretical analyses
\cite{TO,CTEQ,MRS}.
In the CTEQ Collaboration \cite{CTEQ} a number of
theorists, phenomenologists and experimentalists have combined
their efforts in order to
deal properly with all experimental and theoretical
subtleties of quantitative QCD analysis of vast amount of data from
various experiments and processes. In \cite{CTEQ,MRS}
p.d.f. are determined with high
accuracy, unheard of just a few years ago. In such circumstances
a careful reanalysis of various
theoretical uncertainties is clearly needed.
Although most of such uncertainties
are discussed in sufficient detail in review papers like
\cite{TO,Levy}, there is one which has
not been so far satisfactorily covered in either these or in any
other paper I am aware of. It concerns the factorization scheme
(FS) dependence of finite order QCD predictions
in processes involving hadrons in the initial state. The
treatment of this ambiguity presented in
\cite{TO,Levy,Tung} is incomplete and moreover contains an error in
the very central point of the factorization mechanism. As the FS
dependence of p.d.f.  has so far obtained much less attention than it
probably deserves, I discuss in this note several of its
aspects, drawing on analogy with the much more publicised case of
the renormalization scheme (RS) ambiguity of the running coupling
constant (couplant) $\alpha_s$. I think that much of the confusion
and misunderstanding that surrrounds this topic stems from the failure
to distinguish these related but in principle separate uncertainties.

The organization of this paper is as follows. In Section 2 the
notation is introduced and some basic facts about the RS
dependence of the couplant
are recalled. The crucial point, i.e. the dependence of
p.d.f. on the choice of the FS is discussed in Section 3, followed in
Section 4 by a few critical remarks on the currently used p.d.f..
In Section 5 the merits of the so-called ``zero'' FS are discussed,
and in particular it is shown
how it can be used for the construction of
consistent next-to-leading order (NLO) event generators.
The results are summarized and conclusions drawn in Section 6.

\section{Remarks on renormalization scheme dependence}
Before coming to the ambiguity in the definition of p.d.f.
let me recall a few basic
facts about RS dependence of the renormalized
couplant $a(\mu)\equiv g^2(\mu)/4\pi^2$.
In massles QCD (to which I restrict my attention) it obeys
the equation
\begin{equation}
\frac{{\mbox{d}}a(\mu)}{{\mbox{d}}\ln\mu}=
-ba^2\left(1+c_1a(\mu)+
c_2a^2(\mu)+\cdots\right),
\label {da}
\end{equation}
where the coefficients $b,c_1$, are fixed by the number
of quark flavours, while all the higher ones
are essentially free, defining the so called {\it renormalization
convention} (RC), RC=$\{c_i,i\geq2\}$.

In the simplest case this couplant enters the perturbation expansion
of a physical quantity $R$, depending on a single external momemtum
$Q$, in the form
\begin{equation}
R(Q)=a(\mu)\left(1+r_{1}(\mu/Q,c_1)a(\mu)+
r_{2}(\mu/Q,c_1,c_2)a^{2}(\mu)+\cdots\right).
\label {r}
\end{equation}
Although not written out
explicitly, also the couplant $a(\mu)$
(when (\ref{da}) is considered to the k-th order)
depends on all $c_i,i\leq k-2$.
Moreover, both the couplant and the coefficients
$r_k$ depend also on the
specification which of the infinite number of solutions to (\ref{da})
we have in mind. Each of these solutions can be labelled, for instance,
by the familiar $\Lambda$ parameter.
 Combining this last information with that on
$c_i$ defines what is usually called the renormalization scheme.
Only if this RS is fixed does the specification of the scale $\mu$
(together with $c_i,i\geq 2$)
uniquely determine both the couplant and the coefficients $r_k$.
Although I prefer the terminology advocated in \cite{PMS}, where the
the term ``RS'' is reserved for a unique specification of both
$a$ and $r_k$ (for detailed discussion of this point, see \cite{my}),
I adopt in the following the more conventional notation in order to
stay in close contact with \cite{TO,CTEQ,MRS}. In this notation
$\mu$ is set equal to some ``natural'' scale
in the problem and the variation of the RS is parametrized by means
of the corresponding $\Lambda_{\rm RS}$ and the
coefficients $c_i,i\geq 2$.
Considering now (\ref{da},\ref{r}) to the NLO, we are left
with only one degree of freedom, corresponding to the variation
of $\Lambda_{\rm RS}$.
To make the following considerations as transparent as possible,
let me set $c_1=0$ in the rest of this note.
\footnote{No essential conclusion obtained in the
following does, however, depend on this purely technical
simplification.}
The internal consistency of perturbation
theory implies, using (\ref{da}),
the following relation between $\mu, a$ and
$r_1$:
\begin{equation}
r_1(\mu/Q,{\rm RS})=b\ln\left(\frac{\mu}
{\Lambda_{\rm RS}}\right)-\rho=
\frac{1}{a}-\rho\;\;\Rightarrow a=\frac{1}{r_1+\rho}
\label{a(r)}
\end{equation}
where $\rho$
\begin{equation}
\rho=b\ln\left(\frac{Q}
{\Lambda_{\rm RS}}\right)-r_1(\mu=Q,{\mbox{RS}})
\label{rho}
\end{equation}
is the renormalization group (RG) invariant \cite{PMS},
which contains the $Q$-dependence of $R(Q)$. As the variations of
$\mu$ and the RS are actually two sides of the same coin, it is
redundant to vary both of them.
Without the los of generality we can fix $\mu$ (for instance by
setting it equal to some external momentum, like $\mu=Q$) in
(\ref{a(r)}) and elsewhere and vary the RS only.
On the other hand, as the ``natural'' scale in the problem is usually
not so unambiguously defined after all, many authors vary both the
scale $\mu$ and the RS. Although unnecessary complication, this
procedure is certainly legal.
One has, however, to keep in mind that without the specification
of the RS the choice of $\mu$ doesn't fix either the couplant $a(\mu)$
or the coefficients $r_k$.
In different RS's the same $\mu$ implies
different $a(\mu/\Lambda_{\rm RS})$
and $r_k$ and thus the choice of
the RS is as important as that of $\mu$.
With this in mind let me continue to
label the RS by means of $\Lambda_{\rm RS}$, but keep $\mu$ still
as a free parameter.
Eq. (\ref{a(r)}) furthermore
suggests that instead of $\Lambda_{\rm RS}$, the value of
$r_{\rm 1}$ can equally well serve the purpose of labelling
the various RS. Substituting (\ref{a(r)}) into (\ref{r}) and
truncating it to the NLO we get
\begin{equation}
R^{\rm NLO}(\rho,r_1)=\frac{2r_1+\rho}{(r_1+\rho)^2}
=\frac{1}{\rho} \left(\frac{1+2r_1/\rho}{(1+r_1/\rho)^2}\right)
\label{rnlo}
\end{equation}
as an explicit function of $r_1$ and $\rho$.
The obvious consequence of the nontrivial dependence of $R^{\rm NLO}$
on the RS is that it would be a profound mistake to ``transform'' the
couplant from one RS into another by means of equating the NLO
(in fact any finite order) approximation to (\ref{r})
in two different RS, say RS$^{(1)}$, RS$^{(2)}$,
 i.e. by solving the equation
\begin{equation}
a({\rm RS}^{(1)})(1+r_1({\rm RS}^{(1)})a({\rm RS}^{(1)}))=
a({\rm RS}^{(2)})(1+r_1({\rm RS}^{(2)})a({\rm RS}^{(2)}))
\label{error}
\end{equation}
Assuming (\ref{a(r)}) in RS$^{(1)}$ and
expressing $r_1({\rm RS}^{(2)})$
in terms of $a^{(1)}=a({\rm RS}^{(1)}), a^{(2)}=a({\rm RS}^{(2)})$
we get from (\ref{error})
\begin{equation} r_1({\rm
RS}^{(2)})=\frac{2a^{(1)}/a^{(2)}-1}{a^{(2)}}-
\rho\frac{a^{(1)}}{a^{(2)}}
\label{trans}
\end{equation}
which yields the correct relation (\ref{a(r)}) between
$r_1({\rm RS}^{(2)})$ and $a^{(2)}$ only for the trivial case
$a^{(1)}=a^{(2)}$. In other words imposing the relation
 (\ref{error}) leads
to inconsistency for any nontrivial RG
transformation! In the particular case
when RS$^{(2)}$ is defined by means of
the effective charges approach of
\cite{Grunberg}, corresponding to $r_1^{\rm ECH}=0$, (\ref{error})
implies
\begin{equation}
a^{\rm ECH}=\frac{1}{\rho}\left(\frac{1+2r_1^{(1)}/\rho}
{(1+r_1^{(1)}/\rho)^2} \right)
\label{aech}
\end{equation}
while the correct relation reads $a^{\rm ECH}=1/\rho$.

\section{Parton distributions in general factorization scheme}
In processes involving hadrons in the initial state, there is,
beside the mentioned RS dependence of the couplant, another kind
of ambiguity, concerning the definition of p.d.f.
beyond the leading order \cite{Politzer}. Again, to
simplify the discussion as much as possible, I restrict
the discussion to
the nonsinglet (NS) quark distribution functions as revealed in the
lepton-nucleon deep inelastic scattering (DIS). Contrary to the
RS dependence of the couplant \cite{my,Patrick,Pich}, this latter
ambiguity has so far received much less attention \cite{ja}.
This is somewhat surprising, taking into account that its
consequences may actually be even more important than those
discussed in Section 2.

In DIS the factorization theorem \cite{Politzer} implies that the
generic
NS structure function $F^{\rm NS}(x,Q^2)$ can be written
as a convolution (I drop the label ``NS'' in the following)
\begin{equation}
F(x,Q^2)=\int_x^1\frac{dy}{y}q(y,M)
C\left(\frac{Q}{M},\frac{x}{y},a(\mu)\right)
\label{conv}
\end{equation}
of the perturbatively uncalculable quark distribution function
$q(x,M)$, defined at the factorization scale $M$, and obeying the
evolution equation
\begin{equation}
\frac{{\rm d}q(x,M)}{{\rm d}\ln M}=\int_x^1\frac{dy}{y}
\left[a(M)P^{(0)}\left(\frac{x}{y}\right)+a^2(M)P^{(1)}
\left(\frac{x}{y}\right)+\cdots\right]
\label{AP}
\end{equation}
and the hard scattering cross-section $C(Q/M,z,a(\mu))$ admitting
perturbation expansion in powers of the couplant at the hard scattering
scale $\mu$, generally different from $M$
\footnote{If considered to all orders of $a(\mu)$,
$C(Q/M,z,a(\mu))$ doesn't actually depend on $\mu$ \cite{Politzer}.
Contrary to the $M$-dependence of $q(N,M)$ and $C^{(1)}(N,M)$, which is
the basic feature of the factorization theorem and holds to all orders,
the dependence of $C(Q/M,z)$ on
$\mu$ is merely a consequence of truncating expansion (\ref{C})
to a finite order.}
\begin{equation}
C(Q/M,z,a(\mu))=\delta(1-z)+a(\mu)C^{(1)}(Q/M,z)+\cdots
\label{C}
\end{equation}
While $P^{(0)}(z)$ is unique,
both $P^{(1)}$ in (\ref{AP}) and $C^{(1)}$ in (\ref{C})
are ambiguous, but
internal consistency of the factorization procedure links the
variation of the NLO
hard scattering cross-section $C^{(1)}$ with
that of the NLO branching function $P^{(1)}$ \cite{Politzer}:
\begin{equation}
C^{(1)}(Q/M,z,\mbox{RS})=P^{(0)}(z)\ln (Q/M)+P^{(1)}(z)/b+
\kappa (z,{\rm RS})
\label{C1}
\end{equation}
where the FS-invariant function $\kappa(z,{\rm RS})$ still, however,
depends on the RS of the couplant $a(M)$. Similar consistency
conditions do exist at each order of perturbation theory.
In the rest of this note I stay within the NLO approximation.
The dependence of $C^{(1)}$ on the RS of the couplant
appears as a consequence of the fact, that the
r.h.s. of (\ref{AP}) is given as an expansion in
powers of this couplant. I shall return to
this point in the next section.
For fixed RS of the couplant $a(M)$,
$P^{(1)}$, or via (\ref{C1}), $C^{(1)}$,
defines at the NLO the FS of p.d.f. : FS:=\{$P^{(1}(z)$\}.
In the above relations the factorization scale M is again,
as in the case of $\mu$, kept as a free parameter.

The first point I want to emphasize is that the renormalization of the
couplant, to the NLO order fully described by
variations of $\Lambda_{\rm RS}$ in both the couplant $a$ and
the coefficients $r_k$, is
{\it independent} of the FS of p.d.f. as specified by
the NLO branching function $P^{(1)}$ in the evolution equation
(\ref{AP})! {\it Any combination} of the
RS=\{$\Lambda_{\rm RS}$\} and FS=\{$P^{(1)}$\} represents
in principle equally legal definition of p.d.f..  Moreover, as $P^{(1)}$
is a function of $z$, it represents in fact an infinite number of of
degrees of freedom, and thus its variations can be expected to be at
least as important as that of the factorization scale $M$. Also this
point will be elucidated in the next section.

Secondly, let me recall the obvious fact that although
the physical observable $F(x,Q^2)$ is,
when (\ref{AP},\ref{C1}) are taken to all orders,
independent of the parameters
describing the renormalization and factorization schemes,
any finite order approximation to these expansions inevitably leads
to nontrivial dependence of $F(x,Q^2)$ on $M$, RS and FS.
I mention it here as it is related to the basic
question I want to address in this section and which
concerns the way the
p.d.f. transform when the FS=\{$P^{(1)}$\} is varied.
As all the considerations are much more transparent in
terms of conventional moments, defined, for a generic function
$f(x)$, as
\begin{equation}
 f(N)=\int_0^1x^{N-1}f(x)dx
\label{moments}
\end{equation}
let me rewrite (\ref{conv}-\ref{C1}) in terms of them,
explicitly writing out the dependence on both the RS of the couplant
$a(M)$ and the FS of the p.d.f.
($d_N, d_N^{(1)}, \kappa(N)$ are the moments of $P^{(0)}(z),
P^{(1)}(z), \kappa(z)$ respectively):
\begin{equation}
\frac{dq(N,M,{\mbox {RS,FS}})}{d\ln M}=
q(N,M,{\mbox {RS,FS}})\gamma_N; \;\;\;
\gamma_N\equiv d_Na(M,{\mbox {RS}})+
d_N^{(1)}(\mbox{FS})a^2(M,{\mbox {RS}}).
\label{APN}
\end{equation}
This is easily solved:
\begin{equation}
q(N,M,{\mbox {RS,FS}})=
A_N(a(M,{\mbox{RS}}))^{-d_N/b}\exp \left[-a(M,{\mbox{RS}})
d_N^{(1)}({\mbox {FS}})/b\right],
\label{q(N)}
\end{equation}
where the constants $A_N$, introduced in \cite{Politzer},
are independent of $M$ as well as $d^{(1)}_N$.
For moments of the structure function (\ref{conv}) we get, excplicitly
writing out the dependence of the NLO approximation to $F(N,Q^2)$ on M
and FS=\{$d_N^{(1)}$\},:
\begin{equation}
F(N,Q^2,M,{\rm RS},d_N^{(1)})=q(N,M,{\mbox{RS}},d^{(1)}_N)
\left(1+a(\mu,{\mbox{RS}})C^{(1)}(Q/M,N,{\mbox{RS}},d^{(1)}_N)\right),
\label{convmom}
\end{equation}
with the following consistency condition, implied by (\ref{C1}),
\begin{equation} C^{(1)}(Q/M,N,\mbox{RS},d^{(1)}_N)=d_N\ln
\frac{Q}{M}+\frac{d_N^{(1)}}{b}+\kappa(N,{\mbox{RS}}).
\label{C1N}
\end{equation}
For each moment $N$, the
expression (\ref{convmom}) is a function of $M$ which, however,
still depends on
two futher parameters, one
specifying the RS of the couplant and the other ($d^{(1)}_N$)
the FS of p.d.f.. If we now want to transform
$q(N,M,{\mbox{RS}},{\mbox {FS}})$ from one FS=\{$d^{(1)}_N$\} into
another, specified by FS=\{$\overline{d^{(1)}_N}$\}, we again cannot
do so by imposing the relation (the RS-dependence is suppressed in
the rest of this section)
\begin{equation}
q(N,M,d^{(1)}_N)\left(1+a(M)C^{(1)}(Q/M,N,d^{(1)}_N)\right)=
q(N,M,\overline{d^{(1)}_N})\left(1+a(M)C^{(1)}(Q/M,N,
\overline{d^{(1)}_N}\right)
\label{errorn}
\end{equation}
As emphasized above, $F(N,Q^2,M,d_N^{(1)})$ does nontrivially depend
on $M, d_N^{(1)}$, as well as the RS and thus postulating equation
like (\ref{errorn}) would violate this basic feature of finite order
approximations. Formally this is clear from
inserting (\ref{q(N)}) into (\ref{errorn}) and solving the
resulting equation. Similarly to (\ref{trans}), we get the following
relation between
$\overline{C^{(1)}}\equiv C^{(1)}(Q/M,N,\overline{d^{(1)}_N})$ and
$\overline{d^{(1)}_N}$:
\begin{equation}
\overline{d^{(1)}_N}=d^{(1)}_N-\frac{b}{a(M)}\ln \left(
\frac{1+a(M)C^{(1)}}
{1+a(M)\overline{C^{(1)}}}\right)
\label{transd}
\end{equation}
which reduces to the correct one, as given in (\ref{C1}),
only for the trivial case
$\overline{d^{(1)}_N}=d^{(1)}_N$! For any other case
the equation (\ref{transd}) is incompatible with the the consistency
condition (\ref{C1}) and thereby wrong.
I discuss this point in detail as in many papers, including
\cite{TO,Tung,Levy}, the equation
(converted into moments and restricted to the NS channel)
used to transform the
quark distribution function between the so-called DIS and
$\overline{\rm MS}$ ``schemes'' (more on them in the next section)
\begin{equation}
q_{\rm DIS}(N,M)=q_{\overline{\rm MS}}(N,M)
\left(1+a(M)C^{(1)}_{\overline{\rm MS}}(N,M)\right)
\label{chyba}
\end{equation}
is ($C^{(1)}_{\rm DIS}=0$ by definition) precisely of the incorrect
form (\ref{errorn})!

The only theoretically consistent way of transforming
$q(N,M,{\mbox {FS}})$
from one FS=\{$d^{(1)}_N$\} into another is given explicitly
in (\ref{q(N)}) with, as emphasized, the constants $A_N$ held fixed.
In \cite{ja} I have discussed the whole
procedure, based on the use of Jacobi polynomials \cite{Shaw,
jarames}, in $x-$space. Although currently other, superior,
methods of solving the
evolution equations are available \cite{TO}, the fact that Jacobi
polynomials are constructed from conventional moments (\ref{q(N)})
for which we know how the FS transformations operate, makes them
invaluable in this kind of considerations.

Finally a remark. The constants $A_N$ represent the most natural
way of parametrizing the uncalculable
nonperturbative properties of the nucleon. They are not related to
any particular ``initial'' $M_0$, nor to any FS=\{$d^{(1)}_N$\},
but determine the asymptotic
behaviour of $q(N,M,{\mbox{RS}},d^{(1)}_N)$ as $M\rightarrow\infty$,
which is unique.

\section{Remarks on current phenomenology}
In the preceding Section I have discussed the central question of
the FS dependence of the p.d.f.. Let me now turn to the
current phenomenology of DIS, related to this
subject.

The first remark concerns the meaning of the words
``$\overline{\mbox{MS}}$'' and ``DIS'', when used in the connection
with the p.d.f. at the NLO.
The $\overline{\mbox {MS}}$ p.d.f. are defined in \cite{TO} as those
``which appear in the equation such as (\ref{conv}) with hard scattering
part $C^{(1)}$ calculated with the $\overline{\mbox{MS}}$ subtraction
prescription.'' Although correct, this definition is obviously
incomplete, as it specifies merely the RS of the couplant but
tell us nothing about the FS=\{$P^{(1)}(z)\}$
to be used in (\ref{AP})!
Recall that the term $\ln 4\pi-\gamma_{\rm E}$ defining
the $\overline{\mbox{MS}}$ ``subtraction scheme'', is an artifact of
extending the definition of the couplant into 4-$\epsilon$ dimensions.
As any RS of the
couplant may be combined with any FS of p.d.f., there is an
infinite set of $\overline{\mbox{MS}}$-like p.d.f., sharing the same
definition of the couplant, and therefore the right to be called
``$\overline{\mbox{MS}}$'', but arbitrarily differing in $P^{(1)}$.
The one usually understood in the literature under the label
``$\overline{\mbox{MS}}$'' corresponds to the FS used in
\cite{Flor,Curci}, which should be
called ``MS'' as it sets (within the dimensional regularization),
the finite parts of the counterterms, renormalizing the appropriate
composite operators, to zero. Altough the name for
a FS is basically a matter of semantics, the current use of the
term ``$\overline{\mbox{MS}}$'' is misleading as it fails to specify
the FS of the p.d.f. used.

Ambiguity of a different kind is associated with the use of the
label ``DIS''. This scheme is supposed to be defined by the
condition:
\begin{equation}
C^{(1)}_{\rm DIS}(Q=M,z,{\mbox{RS,FS}})=0 \;\;\;\Rightarrow
P^{(1)}_{\rm DIS}(z,{\rm RS})=-b\kappa(z,{\mbox{RS}})
\label{DIS}
\end{equation}
which, however, is again not unique, due to the fact that the FS
invariant $\kappa(z,\mbox{RS})$
still depends on the RS of the couplant. In fact it is the
the combination
\begin{equation}
\varepsilon(z)\equiv\kappa (z,{\mbox{RS}})+
P^{(0)}(z)\ln (Q/\Lambda_{\rm RS})
\label{invariant}
\end{equation}
which is independent of $M$, FS=\{$P^{(1)}(z)\}$ as well as the
RS of the couplant \cite{ja}. Consequently
in a given RS of the couplant $a(M,{\rm RS})$ we find
\begin{equation}
P^{(1)}_{\rm DIS}(z,{\rm RS})=-b\varepsilon(z)+bP^{(0)}(z)\ln
\frac{Q}{\Lambda_{\rm RS}}.
\label{PDIS}
\end{equation}
For example we find
\begin{equation}
P^{(1)}_{\rm DIS}(z,\overline{\mbox{MS}})=
P^{(1)}_{\rm DIS}(z,\mbox{MS})+
bP^{(0)}(z)\ln \frac{\Lambda_{\rm MS}}{\Lambda_{\overline{\rm MS}}}
\label{MSBMS}
\end{equation}
In general, there is again an infinite set of ``DIS''-like p.d.f.,
which share the property $C^{(1)}=0$, but differ in the NLO branching
function $P^{(1)}(z)$ and
the couplant, thereby leading to different numerical
predictions when inserted into (\ref{conv}).  P.d.f. bearing the name
``DIS'', like
those of \cite{DFLM}, or some of \cite{CTEQ}, tacitly assume
$\overline{\mbox{MS}}$ as the RS of the couplant. This, however, is not
a must and thus for an unambiguous specification of the ``DIS''-like FS
the RS of the couplant should always be specified.

The second remark concerns the practical aspect of exploiting
the vast freedom in the definition of p.d.f. at the NLO.
As already mentioned, little phenomenological
attention has so far been payed to the FS dependence of p.d.f..
This may be
due in part to the failure to appreciate the independence of these
two renormalization procedures. It is also true that to apply,
for instance, the idea of ``optimization'' \cite{PMS} to the FC
dependence of p.d.f. in $x$-space is technically much more
involved. In \cite{ja} I have, however, argued that at least
the FS defined by setting $P^{(1)}=0$ (at the NLO; at higher orders
it would generalize by setting all higher order AP branching functions
to zero) should seriously be considered.
In this ``zero'' FS the full NLO correction is put
into the hard scattering cross-section $C^{(1)}$, thereby
representing in some
sense the opposite of the DIS FS, which sets $C^{(1)}(z)=0$.
Moreover, it turns out \cite{ja} that when the moments of p.d.f.
are considered, this FS is very close to that obtained via the
Principle of Minimum Sensitivity of \cite{PMS} (for $c_1=0$ they
even coincide). Although this
results doesn't automatically imply the same close relation for
the p.d.f. themselves, it seems reasonable to add this
FS to the list of those used in phenomenological applications.

\section{``Zero'' FS and NLO event generators}
There is, in fact, another reason why this FS could be
of considerable interest.
Recall, that all currently used event generators, like HERWIG,
PYTHIA, JETSET, LEPTO etc., are based on
essentially leading-log
parton showers. Although they are sometimes combined with NLO
hard scattering cross-sections, the overall description remains
only LO. The simple picture of LO parton
showers becomes much more complicated when one attempts to
generalize them to the NLO. To get a
consistent NLO description of any hard scattering process
in the ``zero'' FS we, however, need merely the LO parton
showers as  $P^{(1)}(z)=0$ in this FS! As the
example of moments of structure functions shows,
this choice may
be quite reasonable and should definitely be tried.
Although simple at first glance, one has to be careful in
taking for the NLO cross-section that corresponding
to this ``zero'' FS. From (\ref{C1}) we easily find
its form:
\begin{equation}
C^{(1)}_{zero}(Q/M,z)\equiv P^{(0)}(z)\ln (Q/M)+\kappa(z,\mbox{RS})
\label{zero}
\end{equation}
Using the results of \cite{Curci} on $P^{(1)}$ and $C^{(1)}$,
$\kappa(z,\mbox{RS})$ can straightforwardly
be evaluated in the $\overline{\mbox{MS}}$ RS. Transformation
to any other RS is then trivial.

Let me stress that this procedure is not equivalent to the
so-called matching of parton showers to fixed order matrix
elements, as recently implemented in LEPTO
event generator \cite{LEPTO}. There, the exact $O(\alpha_s)$
matrix element is matched to the parton shower at some particular
value of incoming parton virtuality $t_m$ (see Fig.1)
in the sense that
below $t_m$ only parton showers are used while above $t_m$ the
matrix element takes fully over.
In the case of ``zero'' FS, the situation is different and the
NLO cross-section ${\rm d}\sigma^{NLO}(Q/M,z,t)/dt$
which, when integrated over $t$, yields
(\ref{zero}) contributes at any virtuality,
even below that given by the factorization scale $t=M^2$,
as only the pole term $1/t$ plus some finite part is factorized
into the parton p.d.f.. As a result, the NLO hard scattering
cross-section ${\rm d}\sigma^{NLO}(Q/M,z,t)/dt$ becomes
a discontinuous function
of $t$ at $t=M^2$, this discontinuity being cancelled by a
similar discontinuity of the parton shower contribution, which
is restricted by definition to the domain $t\leq M^2$!
Work on practical implementation of this idea is in progress.

The situation is sketched in Fig.2a, where the full
${\rm d}\sigma^{NLO}(Q/M,z,t)/dt$, corresponding to the sum of the
diagram in Fig.1 and the one with the gluon radiated off the
outgoing quark, is plotted
as a function of $\tau=-t$ for fixed $Q^2$ and $z$. The full
result is a sum of three terms: the pole term
of the form $A(z)/\tau$,
the residue of which is proportional to $P^{(0)}(z)$, the $\tau$
independent
constant $B(z,Q)$ and $C(z,Q)\tau$, linearly rising with $\tau$.
Notice that for $\tau\rightarrow0$, the finite part of
${\rm d}\sigma^{NLO}(z,Q,t)$,
i.e. the sum of the last two terms,
becomes negative.
The result of factorization, i.e.
the separation of the full NLO cross-section, containing all
infrared and parallel singularities, into a part included
in the quark distribution function and the remaining, finite,
hard scattering cross-section, is
represented in Fig.2b
by the dashed and dotted curves, discontinuous at
the factorization scale $\tau=M^2$.
For $\tau>M^2$ the hard scattering cross-section
${\rm d}\sigma^{NLO}(z,Q,t)$ coincides with the full result,
but for $\tau<M^2$ its definition is ambiguous as it depends on
how much of the finite part will
accompany the sigular pole pole term into the
definition of the quark distribution function. Fig.2b corresponds
to the case that only the pole term $A/\tau$ is subtracted.
Recall that, for instance, the term $P^{(0)}(z)\ln(Q/M)$ appearing
in (\ref{C1}) is essentially the integral of the pole term $A/\tau$
from $M^2$ to the upper kinematically allowed value of $\tau$,
proportional to $Q^2$.

\section{Summary and conclusions}
In this note I have discussed several aspects of factorization
scheme dependence of parton distribution functions. I have
emphasized potential importance of proper treatment of this
ambiguity for the reliablity of theoretical analyses of ever
better data. Special attention has been payed to the correct
transformation of p.d.f. between different factorization schemes
and the ambiguities in the meaning of some of the currently
most popular definition of p.d.f. have been brought to light.
Finally the so-called ``zero'' FS has been proposed and shown
to be potentialy useful in the construction of NLO event
generators.

\parindent 0.01cm
\vspace{0.6cm}
{\Large \bf Acknowledgment}

\vspace{0.4cm}
I am grateful to P. Kol\'{a}\v{r} for careful reading of the
manuscript amd many stimulating discussions.

\vspace{0.6cm}
\parindent 0.01cm
{\Large \bf Figure captions} \\

\vspace{0.3cm}
Fig.1: Feynman diagram describing the process
e$^-$+q$\rightarrow$e$^-$+q+g
with parallel singularity in the $t$-channel.

\vspace{0.3cm}
Fig.2a: A typical shape of ${\rm d}\sigma(Q,z,t)/dt$ as a function
of $\tau=-t$ for fixed $x,Q^2$. In this example the pole term (dashed
curve) is added to a linearly rising finite part (dotted line) to give
the full NLO contribution (solid curve).

\vspace{0.3cm}
Fig.2b: Separation of the full NLO contribution (solid curve)
into the part absorbed in the quark distribution function
(dashed curve) and the finite NLO hard scattering cross-section
(dotted curve). The discontinuities of the last two curves at
$\tau=M^2$ cancel in the sum.
\end{document}